\input harvmac
\def\half{{1 \over 2}}

\def\>{{\rangle}}
\def\<{{\langle}}

\def\p{{\partial}}
\def\s{{\sigma}}

\def\L{{\Lambda}}
\def\O{{\Omega}}

\def\a {{\alpha}}
\def\b {{\beta}}
\def\ad {{\dot \a}}
\def\bd {{\dot \b}}

\def\e {{\epsilon}}

\def \t {{\theta}}
\def \tb {{\bar\theta}}

\Title{\vbox{\hbox{IFT-P.042/96}}}
{\vbox{\centerline{\bf 
Super-Maxwell Actions with Manifest Duality}}}
\bigskip\centerline{Nathan Berkovits}
\bigskip\centerline{Instituto
de F\'{\i}sica Te\'orica, Univ. Estadual Paulista}
\centerline{Rua Pamplona 145, S\~ao Paulo, SP 01405-900, BRASIL}
\bigskip\centerline{e-mail: nberkovi@power.ift.unesp.br}
\vskip .2in
Superstring field theory was recently used to derive a four-dimensional
Maxwell action with manifest duality. This action is related to the 
McClain-Wu-Yu Hamiltonian and can be locally coupled to electric and magnetic
sources. 

In this letter, the manifestly dual Maxwell action is supersymmetrized 
using N=1 and N=2 superspace. The N=2 version may be useful for studying
Seiberg-Witten duality.

\Date{October 1996}
\newsec {Introduction}

The study of N=2 super-Yang-Mills theory has recently become popular due 
to the duality conjecture of Seiberg and Witten.\ref\seiberg{
N. Seiberg and E. Witten, Nucl. Phys. B426 (1994) 19.} This duality
conjecture asserts that by exchanging electrically and magnetically
charged states, the N=2 quantum theory at strong and weak coupling can
be related.

This quantum duality conjecture is related to a continuous classical
symmetry of N=2 super-Maxwell theory which rotates the electric field
into the magnetic field and vice versa. In the standard super-Maxwell
action, classical duality symmetry is not manifest since it rotates
Bianchi identities into equations of motion. This is related to the fact
that magnetic sources cannot couple locally in the standard action.

Recently, superstring field theory was used to derive a ten-dimensional
action for a self-dual five-form field strength.\ref\meone{
N. Berkovits,
``Manifest electromagnetic duality in closed superstring field theory'',
hep-th 9607070, to appear in Phys. Lett. B.} After dimensional reduction,
this gave a manifestly dual action for a four-dimensional Maxwell field.
The action contains an infinite number of fields and can be obtained
from the McClain-Wu-Yu Hamiltonian\ref\MWY 
{B. McClain, Y.S. Wu, and
F. Yu, Nucl. Phys. B343 (1990) 689\semi
I. Martin and A. Restuccia, Phys. Lett. B323 (1994) 311\semi
F.P. Devecchi and M. Henneaux, ``Covariant path
integral for chiral p-forms'', hep-th 9603031.}
by performing a Legendre transformation.\ref\bengtsson
{I. Bengtsson and A. Kleppe, ``On chiral P-forms'', hep-th 9609102.}  
Since duality is manifest, it is easy to couple locally to electric and
magnetic sources.\ref\metwo{N. Berkovits, ``Local actions with
electric and magnetic sources'', hep-th 9610134.}

In this letter, the manifestly dual Maxwell action is generalized
to super-Maxwell actions with N=1 or N=2 four-dimensional supersymmetry.
These actions can couple locally to electric and magnetic
sources and are written in N=1 or N=2 superspace. The N=2 version may
be useful for studying Seiberg-Witten duality since, at the classical
level, duality is now manifest.

In section II of this letter, the manifestly dual Maxwell action is
reviewed. In section III, the action is supersymmetrized in N=1 superspace,
and in section IV, in N=2 superspace. In section V, some conclusions are
presented.

\newsec {Review of the manifestly dual Maxwell action}

In references \meone and \metwo, superstring field theory in the
presence of D-branes was used to obtain a ten-dimensional action
for a self-dual five-form field strength in the presence of sources.
After dimensional reduction to four dimensions, this gave a manifestly
dual Maxwell action in the presence of electric and magnetic sources.
This action is constructed from a complex vector field, $E_p=A_p +iB_p$,
and an infinite set of real anti-symmetric tensor fields,
$F_{(n)}^{pq}$, where $(n)$ runs from 0 to $\infty$. In the presence
of a fermionic dyon source with electric charge $e$ and magnetic
charge $g$, the action is
\eqn\fouraction{{\cal S}=
 \int d^{4}x
[-F_{(0)}^{pq}(\p_p A_q -\half\epsilon_{pqrs}\p^r B^s) }
$$+\half F_{(1)}^{pq}(\p_p A_q +\half\epsilon_{pqrs}\p^r B^s) 
-{1\over 2}
\sum_{n=0}^\infty 
(F_{(2n)}^{pq}
+F_{(2n+2)}^{pq})
F_{(2n+1)pq}$$
$$+
\bar\psi^\ad (i\p_p + e A_p +g B_p) \sigma^p_{\a\ad} \psi^\a]$$
where $\a$ and $\ad$ are two-component Weyl spinor indices,
$\s^0$ is the identity matrix, and $\s^i$ are the Pauli matrices
for $i=1$ to 3. Note that \fouraction can also be written as
\eqn\zeroaction{{\cal S}=
 \int d^{4}x [-{1\over 4} (F_{(0)}^{\a\b} \p_{\a\ad} \bar E^\ad_\b
+ \bar F_{(0)}^{\ad\bd} \p_{\a\ad} E^\a_\bd)}
$$ +{1\over 8} (F_{(1)}^{\a\b} \p_{\a\ad} E^\ad_\b
+ \bar F_{(1)}^{\ad\bd} \p_{\a\ad}\bar E^\a_\bd)$$
$$-{1\over 4}\sum_{n=0}^\infty
((F_{(2n)}^{\a\b} +
F_{(2n+2)}^{\a\b} )
F_{(2n+1)\a\b} 
+(\bar F_{(2n)}^{\ad\bd} +
\bar F_{(2n+2)}^{\ad\bd} )
\bar F_{(2n+1)\ad\bd})$$ 
$$+\bar\psi^\ad (i\p_{\a\ad} +\half (e-ig)E_{\a\ad}
+\half (e+ig) \bar E_{\ad\a})\psi^\a]$$
where $F_{(n)}^{\a\b}=\half \s_p^{\a\ad} \bar\s_{q\ad}^\b F^{pq}_{(n)},$
$E_{\a\ad}=(A_p+iB_p)\s^p_{\a\ad}$, $\bar\s^{p\a\ad}=\e^{\a\b}\e^{\ad\bd}
\s^p_{\b\bd}$, and $\p_{\a\ad}=\s^p_{\a\ad} \p_p.$

The above action is manifestly invariant under the continuous duality
rotation
\eqn\zerodual{F^{\a\b}_{(2n)} \to e^{i\kappa} F^{\a\b}_{(2n)},\quad
F^{\a\b}_{(2n+1)} \to e^{-i\kappa} F^{\a\b}_{(2n+1)},}
$$E^p \to e^{i\kappa} E^p, \quad (e+ig) \to e^{i\kappa} (e+ig)$$
where $\kappa$ is a real constant. It is also invariant under the local
gauge transformation
\eqn\zerogauge{E_p \to E_p + 2 \p_p \L,\quad
\psi^\a \to e^{(ie+g)\L +(ie-g)\bar\L} \psi^\a .}

Since an infinite number of fields are present in \fouraction, there
are subtleties involved in analyzing solutions to the equations of motion.
To remove these subtleties, only solutions with a finite number of
non-zero fields will be allowed. In other words, at each point in 
spacetime, only a finite number of on-shell fields will be allowed to
be non-zero. This restriction is similar to that of reference \bengtsson
and guarantees that the energy is finite and well-defined.\foot{
In reference \bengtsson, the fields with label $(n)$ were restricted to
satisfy $|F_{(n)}|< 1/n$ for $n>N$. This type of restriction is
inappropriate for fermionic fields.}

The equations of motion from varying $F^{pq}_{(n)}$, $A_p$, $B_p$, and
$\psi^\a$ are easily calculated to be
\eqn\zeroempre{
F_{(0)}^{pq} -\half(\p^p A^q -\p^q A^p +\e^{pqrs} \p_r B_s) 
= -F_{(2)}^{pq}
= F_{(4)}^{pq}
=- F_{(6)}^{pq}= ... ,}
$$\p^p A^q -\p^q A^p -\e^{pqrs} \p_r B_s 
= F_{(1)}^{pq}
= -F_{(3)}^{pq}
= F_{(5)}^{pq}= ... ,$$
$$\p_q F^{pq}_{(0)}= e \bar\psi^\ad \sigma^p_{\a\ad}\psi^\a,\quad
\half\epsilon_{pqrs}\p^q F^{rs}_{(0)}= g
\bar\psi^\ad \sigma^p_{\a\ad}\psi^\a.$$
$$(\p_p -i e A_p -i g B_p)\s^p_{\a\ad} \psi^\a =0.$$

Solutions to \zeroempre with a finite number of non-zero fields satisfy
\eqn\zeroem{
F_{(0)}^{pq} =\p^p A^q -\p^q A^p =\e^{pqrs} \p_r B_s, \quad 
 F_{(n+1)}^{pq}=
 F_{(n+2)}^{pq}= 0 ,}
$$\p_q F^{pq}_{(0)}= e \bar\psi^\ad \sigma^p_{\a\ad}\psi^\a,\quad
\half\epsilon_{pqrs}\p^q F^{rs}_{(0)}= g
\bar\psi^\ad \sigma^p_{\a\ad}\psi^\a.$$
$$(\p_p -i e A_p -i g B_p)\s^p_{\a\ad} \psi^\a =0.$$
These are the standard Maxwell equations in the presence of electric
and magnetic sources.

\newsec {Manifestly dual N=1 super-Maxwell action}

To generalize \zeroaction to N=1 superspace, one introduces
a complex scalar superfield, $V$ and $\bar V$,
and an infinite set of chiral spinor superfields,
$W_{(n)}^{\a}$
and $\bar W_{(n)}^{\ad}$, satisfying $\bar D_\ad W^\b_{(n)}=$
$ D_\a \bar W^\bd_{(n)}=0.$ ($D_\a = \p /\p \t^\a +i\tb^\ad\p_{\a\ad}$ and
$\bar D_\ad = \p /\p \tb^\ad +i\t^\a\p_{\a\ad}$ are the usual N=1
fermionic derivatives.) This is the N=1 analog of the fields in
\fouraction since the standard N=1 super-Maxwell action uses a
real
scalar 
superfield whose field strength is a chiral spinor superfield.\ref
\Salam{A. Salam and B. Strathdee, Phys. Rev. D11 (1975) 1521.}

The N=1 super-Maxwell action in the presence of a dyonic Wess-Zumino
scalar multiplet\ref\WZ{J. Wess and B. Zumino, Phys. Lett. B49 (1974)
52.} is
\eqn\oneaction{{\cal S}=
\int d^{4}x d^2 \t d^2\tb
[-{1\over 8} (W_{(0)}^{\a} D_{\a} \bar V
+ \bar W_{(0)\ad} \bar D^\ad V)}
$$  +{1\over {16}} (W_{(1)}^{\a} D_{\a} V
+ \bar W_{(1)\ad} \bar D^\ad \bar V)]$$
$$-{1\over 4}\int d^4 x d^2 \t \sum_{n=0}^\infty
(W_{(2n)}^\a +
W_{(2n+2)}^\a )
W_{(2n+1)\a}$$ 
$$-{1\over 4}\int d^4 x d^2 \tb \sum_{n=0}^\infty
(\bar W_{(2n)\ad} +
\bar W_{(2n+2)\ad} )
\bar W_{(2n+1)}^\ad $$
$$+\half\int d^4 x d^2\t d^2\tb
~\bar\Phi e^{(e-ig)V
+(e+ig) \bar V}\Phi$$
where $\Phi$ is a chiral scalar superfield satisfying $\bar D^\ad \Phi=0.$

This action is manifestly invariant under the duality
rotation
\eqn\onedual{W^{\a}_{(2n)} \to e^{i\kappa} W^{\a}_{(2n)},\quad
W^{\a}_{(2n+1)} \to e^{-i\kappa} W^{\a}_{(2n+1)},}
$$V \to e^{i\kappa} V, \quad (e+ig) \to e^{i\kappa} (e+ig),$$
and under the
gauge transformation
\eqn\onegauge{V \to V + (D)^2 \L +(\bar D)^2 \bar\O,\quad
\Phi \to e^{(ig-e)(\bar D)^2\bar\O -(ig+e)(\bar D)^2\bar\L} \Phi ,}
where $(D)^2 =\half D^\a D_\a$.

Assuming that only a finite number of on-shell fields are non-zero,
the equations of motion for \oneaction are
\eqn\oneem{
W_{(0)}^{\a} ={1\over 4}(\bar D)^2 D^\a (V+\bar V)=
{1\over 4}(\bar D)^2 D^\a (V-\bar V),\quad
 W_{(n+1)}^{\a}=
 W_{(n+2)}^{\a}= 0 ,}
$${1\over 4}D_\a W^\a_{(0)}= (e+ig) \bar\Phi 
e^{(e-ig)V
+(e+ig) \bar V}\Phi,$$
$$(D)^2 (e^{(e-ig)V
+(e+ig) \bar V}\Phi)=0.$$
These equations are easily seen to be the N=1
generalization of \zeroem where
$F_{(n)}^{\a\b}={i \over 2}(D^\a W_{(n)}^\b
+D^\b W_{(n)}^\a)|_{\t=\tb=0},$, 
$E_{\a\ad}= D_\a \bar D_\ad V|_{\t=\tb=0}$, 
and
$\psi_{\a}= D_\a\Phi|_{\t=\tb=0}$.

\newsec {Manifestly dual N=2 super-Maxwell action}

To generalize \zeroaction to N=2 superspace, one introduces
a complex SU(2) triplet of superfields, $V_{jk}$ and $\bar V_{jk}$,
and an infinite set of chiral scalar superfields,
$W_{(n)}$
and $\bar W_{(n)}$, satisfying $\bar D_\ad^j W_{(n)}=$
$ D_\a^j \bar W_{(n)}=0.$ ($D_\a^j = \p /\p \t^\a_j +i\tb^{j\ad}
\p_{\a\ad}$ and
$\bar D_\ad^j = \p /\p \tb^\ad_j +i\t^{j\a}\p_{\a\ad}$ are the usual N=2
fermionic derivatives where $j$ is an
SU(2) doublet index.) This is the N=2 analog of the fields in
\fouraction since the standard N=2 super-Maxwell action uses a
real
triplet of 
superfields whose field strength is a chiral scalar superfield.\ref
\Grimm{R. Grimm, M. Sohnius and J. Wess, Nucl. Phys. B133 (1978) 275.}

The N=2 super-Maxwell action in the presence of a dyonic Fayet-Sohnius
scalar hypermultiplet\ref\FS{P. Fayet, Nucl. Phys. B113 (1976) 135\semi
M. Sohnius, Nucl. Phys. B138 (1978) 109.}
is
\eqn\twoaction{{\cal S}=
\int d^{4}x d^4 \t d^4\tb
[-{1\over {48}} (W_{(0)} D^{j\a} D^k_{\a} \bar V_{jk}
+ \bar W_{(0)}\bar D^j_\ad \bar D^{k\ad} V_{jk})}
$$+{1\over {96}} (W_{(1)} D^{j\a} D^k_{\a}  V_{jk}
+ \bar W_{(1)}\bar D^j_\ad \bar D^{k\ad} \bar V_{jk})]$$
$$-{1\over 4}\int d^4 x d^4 \t \sum_{n=0}^\infty
(W_{(2n)} +
W_{(2n+2)} )
W_{(2n+1)}$$ 
$$-{1\over 4}\int d^4 x d^4 \tb \sum_{n=0}^\infty
(\bar W_{(2n)} +
\bar W_{(2n+2)} )
\bar W_{(2n+1)} $$
$$+\half\int d^4 x d^2\t^+ d^2\tb^+\int du
(\overline{\Phi^+})^* (D^{++}+(e-ig) V^{++} +
(e+ig)(\overline{V^{++}})^*) \Phi^+$$
where $V^{++}= (D^+)^2 (\bar D^+)^2 u_j^- u_k^- V^{jk}$,
$(\overline{V^{++}})^*=
(D^+)^2 (\bar D^+)^2 u_j^- u_k^-\bar V^{jk}$,
$D^\pm_\a=u^\pm_j D_\a^j$,
$\bar D^\pm_\ad=u^\pm_j \bar D_\ad^j$,
$D^{++}= u_j^+ \p/\p u_j^-$, and
$\Phi^+$ is an analytic superfield satisfying $D_\a^+ \Phi^+$=
$\bar D^+_\ad \Phi^+=0$.

As discussed in reference \ref\harm{A. Galperin, E. Ivanov, S. Kalitzin,
V. Ogievetsky and E. Sokatchev,  
Class. Quant. Grav. 1 (1984) 469.}, $u_j^\pm$ are harmonic variables
which are needed for coupling a hypermultiplet in N=2 superspace.
They are complex variables satisfying $u_j^+ u^{j+}$=
$u_j^- u^{j-}$=0 and 
$u_j^+ u^{j-}$=1. The bar operation acts on all fields as complex conjugation,
while the $*$ operation acts only on the $u_j^\pm$ variables as
$(u_j^\pm)^*=\pm u_j^\mp$. Note that 
$(\overline{u^{j+}})^*=u_j^+$, so 
$D^+_\a (\overline{\Phi^+})^*=$
$\bar D^+_\ad (\overline{\Phi^+})^*=0.$
Therefore, one only needs to integrate the source term over
$d^2 \t d^2\tb$ since it is annihilated by $D_\a^+$ and $\bar D_\ad^+$.
Integration over the $u_j^\pm$ variables is defined by $\int du ~1 =1$ and
$\int du~u^+_{(j_1} ... u^+_{j_M}
u^-_{k_1} ... u^-_{k_N)} =0.$ For more details on harmonic superspace
notation, see reference \harm.

The N=2 action of \twoaction is manifestly invariant under the duality
rotation
\eqn\twodual{W_{(2n)} \to e^{i\kappa} W_{(2n)},\quad
W_{(2n+1)} \to e^{-i\kappa} W_{(2n+1)},}
$$V_{jk} \to e^{i\kappa} V_{jk}, \quad (e+ig) \to e^{i\kappa} (e+ig),$$
and under the
gauge transformation
\eqn\twogauge{V_{jk} \to V_{jk} +3 D^{l\a} \L_{jkl\a} +3\bar D^l_\ad
\bar\O_{jkl}^\ad,$$
$$
\Phi^+
\to e^{(D^+)^2 (\bar D^+)^2[(ig-e)f_{++++} -(ig+e)(\overline{f_{++++}})^*]}
\Phi^+,}
where 
$\L_{jkl}^\a$ and $\O_{jkl}^\ad$ are complex SU(2) quadruplets,
$f_{++++}=$
$u_+^j u_+^k u_+^l u_+^m$
$ (D_j^\a \L_{klm\a}+
\bar D_{j\ad}$
$\bar\O_{jkl}^\ad),$ and
$(\overline{f_{++++}})^*=$
$u_+^j u_+^k u_+^l u_+^m $
$(\bar D_{j\ad}\bar \L_{klm}^\ad+$
$ D_j^\a\O_{jkl\a}).$

Assuming that only a finite number of on-shell fields are non-zero,
the equations of motion for \twoaction are
\eqn\twoem{
W_{(0)} ={1\over {24}}(\bar D)^4 D^{j\a} D^k_\a (V_{jk}+\bar V_{jk})=
{1\over {24}}(\bar D)^4 D^{j\a} D^k_\a (V_{jk}-\bar V_{jk}),}
 $$W_{(n+1)}=
 W_{(n+2)}= 0 ,$$
$${1\over {24}}D^{j\a} D_\a^k W_{(0)}= (e+ig)\int du (\overline{\Phi^+})^* 
u_+^j u_+^k \Phi^+,$$
$$(D^{++} +(e-ig) V^{++} +(e+ig) (\overline{V^{++}})^*) \Phi^+ =0.$$
These equations are easily seen to be the N=2
generalization of \zeroem where
$F_{(n)}^{\a\b}$
=${1 \over 2}D_j^\a D^{j\b} W_{(n)}|_{\t=\tb=0}$, 
$E^{\a\ad}= {1\over {36}}$
$(D_j^\a D^{j\b} D_{l\b})$
$(\bar D_k^\ad \bar D^k_\bd \bar D_m^\bd)$
$ V^{lm}|_{\t=\tb=0}$, 
and
$\psi^{\a}=\int du D_+^\a\Phi^+|_{\t=\tb=0}$.

\newsec {Conclusions}

In this letter, N=1 and N=2 super-Maxwell actions were constructed 
with manifest duality. Since these actions can be coupled locally to
electric and magnetic sources, the N=2 version may be useful for
studying Seiberg-Witten duality.

The original form for the manifestly dual Maxwell action was found
by computing the contribution of massless Ramond-Ramond fields in
the closed superstring field theory action.\meone
Since this computation used the Ramond-Neveu-Schwarz formalism of the
superstring (including some extra non-minimal fields\ref\sieg
{W. Siegel, Int. J. Mod. Phys. A6 (1991) 3997\semi
N. Berkovits, M.T. Hatsuda and W. Siegel, Nucl. Phys. B371
(1991) 434.}), the result was not manifestly spacetime-supersymmetric.
For describing four-dimensional compactifications of the Type II superstring,
there also exists a manifestly N=2 super-Poincar\'e invariant formalism.\ref
\mefour{N. Berkovits, Nucl. Phys. B431 (1994) 258.}\ref\mesieg{ 
N. Berkovits and W. Siegel, Nucl. Phys. B462 (1996) 213.}
It would be interesting to try to directly derive the N=2 super-Maxwell 
action of \twoaction using superstring field theory in the 
super-Poincar\'e invariant formalism. This may be possible since,
unlike the standard N=2 super-Maxwell action, \twoaction does not
require restricted chiral superfields satisfying
$D^{j\a} D^k_\a W=\bar D_\ad^j \bar D^{k\ad} \bar W$. As was
discussed in \mesieg, restricted chiral superfields are unnatural in
the super-Poincar\'e invariant formalism since there is no
two-dimensional analog of restricted chirality.

\vskip 20pt

{\bf Acknowledgements:} I would like to thank Warren Siegel
for useful discussions. This
work was financially supported by the Brazilian 
FAPESP.

\listrefs
\end